\documentclass[graybox]{svmult}

\usepackage{type1cm}        % activate if the above 3 fonts are
\usepackage{makeidx}         % allows index generation
\usepackage{graphicx}        % standard LaTeX graphics tool
\usepackage{multicol}        % used for the two-column index
\usepackage[bottom]{footmisc}% places footnotes at page bottom

\usepackage{cite}
\usepackage{epsfig}
\usepackage{graphicx}
\usepackage{amsmath}
\usepackage{amssymb}
\usepackage{ulem}
\usepackage{mdwlist}
\usepackage{color}
\usepackage{color}
\usepackage{slashed}

\makeindex             % used for the subject index
\newcommand{\Mvec}{{\rm\bf M}}
\newcommand{\ep}{\varepsilon}
\newcommand{\Li}{{\rm Li}}
\newcommand{\HA}{{\rm H}}

\begin{document}

\title*{{\footnotesize{\sf DESY 18--216,~~DO-TH 18/28}}\\Large scale analytic calculations in quantum field theories}
\titlerunning{Large scale analytic calculations in QFT} 
% your contribution title if the original one is too long
\author{Johannes Bl\"umlein}
% Use \authorrunning{Short Title} for an abbreviated version of
% your contribution title if the original one is too long
\institute{Deutsches Elektronen-Synchrotron, DESY, Platanenallee 6, D-15738 Zeuthen, Germany, 
\email{Johannes.Bluemlein@desy.de}}
%
% Use the package "url.sty" to avoid
% problems with special characters
% used in your e-mail or web address
%
\maketitle

%%%%%%%%%%%%%%%%%%%%%%%%%%%%%%%%%%%%%%%%%%%%%%%%%%%%%%%%%%%%%%%%%%%%%%%%%%%%%%%%%%%%%%%%%%%%%%%
\abstract{
We present a survey on the mathematical structure of zero- and single scale quantities and the 
associated calculation methods and function spaces in higher order perturbative calculations 
in relativistic renormalizable quantum field theories.}
%%%%%%%%%%%%%%%%%%%%%%%%%%%%%%%%%%%%%%%%%%%%%%%%%%%%%%%%%%%%%%%%%%%%%%%%%%%%%%%%%%%%%%%%%%%%%%%

%%%%%%%%%%%%%%%%%%%%%%%%%%%%%%%%%%%%%%%%%%%%%%%%%%%%%%%%%%%%%%%%%%%%%%%%%%%%%%%%%%%%%%%%%%%%%%%
\section{Introduction}
\label{sec:1}
%%%%%%%%%%%%%%%%%%%%%%%%%%%%%%%%%%%%%%%%%%%%%%%%%%%%%%%%%%%%%%%%%%%%%%%%%%%%%%%%%%%%%%%%%%%%%%%

\vspace{1mm}\noindent
Precise theoretical predictions within the Standard Model of elementary particles are indispensable
for the concise understanding of the fundamental parameters of this physical theory and the 
discovery of its potential extensions. At the experimental side highly precise measurements exist
at $e^+e^-$, $ep$ and $pp$--colliders as at LEP, HERA, and the LHC. In the near future the high luminosity
phase of the LHC will even provide much more precise data. Other facilities, like the ILC \cite{ILC} 
and a possible FCC \cite{FCC}, are currently planned. During the last three decades enormous efforts have been made to 
calculate key observables measured at these colliders at higher and higher accuracy, to meet the challenge
provided by the accuracy of the experiments.

For zero--scale quantities currently analytic massless calculations can be performed at the five--loop 
and for massive calculations at the four-loop level. Single scale calculations are performed in both cases
at the three--loop level. To perform these large scale calculations very demanding efforts are needed
at the side of their automation, computer--algebraic implementation, and the use of highly efficient
mathematical technologies. Therefore, the present problems can only be solved within a very close 
interdisciplinary cooperation between experts in all these different fields and it cannot be the sole 
tasks for theoretical physicists anymore.

While at one--loop order the mathematical solution for many scattering processes has been known early,
cf.~\cite{tHooft:1978jhc,tHooft:1973wag,Veltman:1994wz}, systematic representations at higher loop order
turned out to be more difficult. The core problem concerns the analytic integration of Feynman parameter 
integrals. Here integration is understood as anti--differentiation. An essential question is to determine 
the final solution space to which the respective integrals do belong and its mathematical structure, and to 
find the irreducible objects through which the corresponding integrals are represented. Furthermore, one 
needs efficient mathematical and computer--algebraic technologies to map the given Feynman parameter 
integrals into the latter quantities.

In this paper we give a survey on the main technological steps to calculate higher loop zero-- and 
single--scale quantities in renormalizable quantum field theories, with the focus on analytic integration
techniques and the occurring function spaces.  The systematic theory of integration in this field is
vastly developing and many more new structures are expected to be revealed in the future at higher loop levels
and by considering the production of more particles in the final state of the respective scattering 
processes. These calculations are needed to obtain stable theoretical predictions for the experimental 
precision measurements at the present and future colliders, which operate at high luminosity.

The paper is organized as follows. In Section~\ref{sec:2} we summarize the main steps in 
multi--loop perturbative calculations. Different methods used in symbolic calculations of
zero-- and single--scale Feynman parameter integrals are described in Section~\ref{sec:3}. In 
Section~\ref{sec:4} a hierarchy of function spaces, mainly for single--scale integrals, is 
discussed  which emerge in present multi--loop calculations. Here we consider as well the representations in
Mellin--$N$ and $x$--space. Section~\ref{sec:5} contains the conclusions.\footnote{
For other recent surveys on integration methods for Feynman integrals see 
\cite{Weinzierl:2010ps,Ablinger:2013eba,Ablinger:2013jta,Weinzierl:2013yn,Duhr:2014woa,Blumlein:2018cms}.}
%%%%%%%%%%%%%%%%%%%%%%%%%%%%%%%%%%%%%%%%%%%%%%%%%%%%%%%%%%%%%%%%%%%%%%%%%%%%%%%%%%%%%%%%%%%%%%%
\section{Main Steps in Multi--Loop Perturbative Calculations}
\label{sec:2}
%%%%%%%%%%%%%%%%%%%%%%%%%%%%%%%%%%%%%%%%%%%%%%%%%%%%%%%%%%%%%%%%%%%%%%%%%%%%%%%%%%%%%%%%%%%%%%%

\vspace{1mm}\noindent
In most of the large projects, which are currently dealt with, the Feynman diagrams are generated 
using packages like {\tt QGRAF} \cite{Nogueira:1991ex} and performing the color algebra for the gauge 
groups using {\tt Color} \cite{vanRitbergen:1998pn}. Standardized algorithms to obtain Feynman parameterizations
exist, cf. e.g. \cite{NAKANISHI,LEFSCHETZ,Bogner:2010kv}. At growing complexity, to perform the Dirac- and spin-algebra will 
be a challenge even to {\tt FORM}  \cite{Vermaseren:2000nd, Tentyukov:2004hz, Tentyukov:2007mu, Ruijl:2017dtg}. 
One further maps the set of the contributing
Feynman integrals to master integrals using the integration--by--parts (IBP) technique \cite{Chetyrkin:1981qh}
based on Laporta's algorithm \cite{Laporta:2001dd}, of which several implementations exist, cf. e.g.
\cite{Smirnov:2008iw,Smirnov:2019qkx,Studerus:2009ye,vonManteuffel:2012np,MARSEID} and others. The remaining 
main step is then the 
integration of the master integrals. One possibility to inspect the problem on hand, is to analyze the associated 
system of first order differential equations for the master integrals.
Sometimes it is also useful to consider, in addition, the related system of linear difference equations.
One may decouple these systems using the algorithms implemented in the packages \cite{ORESYS,BCP13},
as e.g. Z\"urcher's algorithm \cite{Zuercher:94}. This leads to a single differential equation or difference
equation of large order and degree and associated determining equations for the remaining master integrals.
If the former equations can be factored at first order, it is known that the master integrals can be 
obtained in terms of indefinitely nested sums or iterated integrals over certain alphabets, which are 
revealed in the solution process, e.g. using difference field and ring theory \cite{Karr:1981,Bron:00,Schneider:01,
Schneider:04a,Schneider:05a,Schneider:05b,Schneider:07d,Schneider:10b,Schneider:10c,Schneider:15a,Schneider:08c,
Schneider:08d,Schneider:08e}, algorithmically implemented in the package {\tt Sigma} \cite{Schneider:2007a,Schneider:2013a}. 
This applies to a wide class of physical cases. Most of the integration and summation methods described in Section~3
apply to them and allow to obtain the integrals analytically in terms of the mathematical functions described in Section~4. 
Finally, efficient numerical representations of these functions have to be provided to obtain numerical predictions
of the different observables for the experiments.
%%%%%%%%%%%%%%%%%%%%%%%%%%%%%%%%%%%%%%%%%%%%%%%%%%%%%%%%%%%%%%%%%%%%%%%%%%%%%%%%%%%%%%%%%%%%%%%
\section{Symbolic Integration of Feynman Parameter Integrals}
\label{sec:3}
%%%%%%%%%%%%%%%%%%%%%%%%%%%%%%%%%%%%%%%%%%%%%%%%%%%%%%%%%%%%%%%%%%%%%%%%%%%%%%%%%%%%%%%%%%%%%%%

\vspace{1mm}\noindent
In the following we summarize main aspects of the analytic integration of multi--loop Feynman parameter
integrals. Of course these integrals can also be evaluated numerically, without observing their
particular analytic structure, to some accuracy and methods 
exist to separate the different pole contributions in $\ep$, cf. e.g. \cite{Binoth:2000ps,Nagy:2006xy,Anastasiou:2007qb,
Smirnov:2008py,Carter:2010hi,Smirnov:2009pb,Becker:2010ng,Becker:2012aq,Becker:2011vg,Smirnov:2015mct,Borowka:2018dsa}, 
which we will not discuss in the following. These methods play a role, however, also for testing analytic results.
In calculating all the integrals required to solve a large scale problem, it is usually necessary to 
combine different analytic methods, at least for the sake of efficiency.
This requirement finally led to the creation of these methods. In the future even more and further 
refined technologies will be needed to solve more enhanced problems. Finally, one ends up with sets
of irreducible functions which span the solutions, see~Section~\ref{sec:4}. The numerical 
representation of these functions is necessary and will be discussed in Section~\ref{sec:43}.

Non of the different techniques described in the following are of universal character. In particular
the solution of the most advanced problems will need a combined and sensible use of various of them. 
All of them have to be handled with care to achieve a steady stepwise reduction of the problem on hand 
and to avoid to enlarge the complexity, given the limited time and memory resources for the corresponding
computer algebraic calculations. This will also apply to future developments, since more complex 
calculations will require further new and advanced technologies. 

Many of the formalisms described below lead to summation problems. Their solution requires dedicated
and efficient algorithms in difference field theory as implemented in the packages {\tt Sigma} 
\cite{Schneider:2007a,Schneider:2013a}, {\tt EvaluateMultiSums} and {\tt SumProduction}
\cite{Ablinger:2010pb,Blumlein:2012hg,Schneider:2013zna}, see also \cite{Schneider:19}. 
%%%%%%%%%%%%%%%%%%%%%%%%%%%%%%%%%%%%%%%%%%%%%%%%%%%%%%%%%%%%%%%%%%%%%%%%%%%%%%%%%%%%%%%%%%%%%%%
\subsection{The PSLQ Method}
\label{sec:31}
%%%%%%%%%%%%%%%%%%%%%%%%%%%%%%%%%%%%%%%%%%%%%%%%%%%%%%%%%%%%%%%%%%%%%%%%%%%%%%%%%%%%%%%%%%%%%%%

\vspace{1mm}
\noindent
The PSLQ method applies to the solution of zero dimensional quantities, i.e. physical quantities
given by pure numbers. If the pool of constants is known or can be guessed over which the corresponding 
quantity has a polynomial representation over $\mathbb{Q}$, a highly precise numerical representation
of the quantity and the individual monomials allows to determine the corresponding rational coefficients,
cf.~\cite{PSLQ}. This method has been applied recently in a massive calculation of the five--loop QCD 
$\beta$-function \cite{Luthe:2017ttg}. Here the individual master integrals certainly contain also 
constants of elliptic nature and probably beyond. However, they all cancel in the final result, which is 
spanned by multiple zeta values (MZVs) \cite{Borwein:1999js,Blumlein:2009cf}, more precisely by 
$\{\zeta_2, \zeta_3, \zeta_5, \zeta_7 \}$, beyond pure rational terms. Let us illustrate the method by an 
example. We would like to determine the harmonic polylogarithm $H_{-1,0,0,1}(1)$, cf.~Section~\ref{sec:42},
which is given by a polynomial of MZVs up to weight {\sf w=4}. I.e. we have 
to apply the PSLQ method over all monomials up to {\sf w=4}  
%--------------------------------------------------------------------------------------------------------
\begin{eqnarray}
\left\{
\ln(2), \zeta_2, \zeta_3, \Li_4\left(\frac{1}{2}\right) \right\}.
\end{eqnarray}
%--------------------------------------------------------------------------------------------------------
An approximate numerical value of $H_{-1,0,0,1}(1)$ is
%--------------------------------------------------------------------------------------------------------
\begin{align}
&0.3395454690873598695906678484608602061387815339795751791304750 \nonumber\\ 
&222490137419723806082682624196443182167020255697096551752247012 \nonumber\\
&11749559277
\end{align}
%--------------------------------------------------------------------------------------------------------
and PSLQ yields
%--------------------------------------------------------------------------------------------------------
\begin{eqnarray}
H_{-1,0,0,1}(1) = -\frac{1}{12} \ln^4(2) + \frac{1}{2} \ln^2(2) \zeta_2 + \frac{3}{5} \zeta_2^2
-\frac{3}{4} \ln(2) \zeta_3  - 2 \Li_4\left(\frac{1}{2}\right).
\nonumber\\
\end{eqnarray}
%--------------------------------------------------------------------------------------------------------
In particular, monomials like $\ln(2), \ln^2(2), \ln^3(2), \zeta_2, \zeta_3$ do not contribute here.
%%%%%%%%%%%%%%%%%%%%%%%%%%%%%%%%%%%%%%%%%%%%%%%%%%%%%%%%%%%%%%%%%%%%%%%%%%%%%%%%%%%%%%%%%%%%%%%
\subsection{Hypergeometric Functions and their Generalizations}
\label{sec:32}
%%%%%%%%%%%%%%%%%%%%%%%%%%%%%%%%%%%%%%%%%%%%%%%%%%%%%%%%%%%%%%%%%%%%%%%%%%%%%%%%%%%%%%%%%%%%%%%

\vspace{1mm}
\noindent
Simpler Feynman--parameter integrals have representations in terms of generalized hypergeometric functions
\cite{HYPKLEIN,HYPBAILEY,SLATER1} and their generalizations such as Appell-, Kampe-De-Feriet- and related functions
\cite{APPEL1,APPEL2,KAMPE1,EXTON1,EXTON2,SCHLOSSER,Anastasiou:1999ui,Anastasiou:1999cx,SRIKARL,Lauricella:1893,
Saran:1954,Saran:1955}. This is due to the hyperexponential nature of the Feynman--parameter integrals, implying real 
exponents due to the dimensional parameter $\ep$. These representations map multiple integrals to single series 
(for generalized hypergeometric functions) and double infinite series (e.g. for Appell series), which finally have to be 
solved by applying summation theory. The simplest function is Euler's Beta-function implying the series of $_{p+1}F_p$ 
functions
%---------------------------------------------------------------------------------------------------------
\begin{eqnarray} 
B(a_1,a_2)           &=& \int_0^1 dt~t^{a_1-1} (1-t)^{a_2-1}
\\
_3F_2(a_1,a_2,a_3;b_1,b_2;x) &=& \frac{\Gamma(b_2)}{\Gamma(a_3) \Gamma(b_2-a_3)} \int_0^1 dt~t^{a_3-1}
(1-t)^{-a_3+b_2-1}
\nonumber\\ && \times
_2F_1(a_1,a_2;b_1;tx).
\end{eqnarray}
%---------------------------------------------------------------------------------------------------------
Representations of this kind are usually sufficient for massless and massive single--scale two--loop problems 
\cite{Hamberg:1990np,HAMBERG,Buza:1995ie,Bierenbaum:2007qe}. In the case of three--loop ladder graphs Appell-functions 
are appearing \cite{Ablinger:2012qm,Ablinger:2015tua}. There are some more classes of higher transcendental functions 
of this kind, which have been studied in the mathematical literature \cite{EXTON1,EXTON2,SRIKARL}. The corresponding 
representations allow the expansion in the dimensional parameter $\ep$. At a given level in the calculation 
of Feynman diagrams one will not find corresponding known function representations and one has to invoke other methods
of integration. One way to derive analytic infinite sum representations are Mellin--Barnes integrals to which we turn now.
%%%%%%%%%%%%%%%%%%%%%%%%%%%%%%%%%%%%%%%%%%%%%%%%%%%%%%%%%%%%%%%%%%%%%%%%%%%%%%%%%%%%%%%%%%%%%%%
\subsection{Analytic Solutions using Mellin--Barnes Integrals}
\label{sec:33}
%%%%%%%%%%%%%%%%%%%%%%%%%%%%%%%%%%%%%%%%%%%%%%%%%%%%%%%%%%%%%%%%%%%%%%%%%%%%%%%%%%%%%%%%%%%%%%%

\vspace{1mm}
\noindent
The higher transcendental functions discussed in Section~\ref{sec:32} have representations in terms of 
Pochhammer--Umlauf integrals \cite{POCHHAMMER,HYPKLEIN,KF} and related to it, by Mellin--Barnes integrals
\cite{BARNES1,MELLIN1}.
They are defined by 
%---------------------------------------------------------------------------------------------------------
\begin{eqnarray} 
\frac{1}{(a+b)^\alpha} = \frac{1}{\Gamma(\alpha)} \frac{1}{2\pi i} \int_{-i \infty}^{i \infty} dz \Gamma(\alpha + z)
\Gamma(-z) \frac{b^z}{a^{\alpha+z}}, ~~~\alpha \in \mathbb{R}, \alpha > 0,
\end{eqnarray} 
%---------------------------------------------------------------------------------------------------------
cf. e.g.~\cite{SMIRNOV}. Here the contour integral is understood to be either being closed to the left or the right
surrounding the corresponding singularities.
The Mellin--Barnes decomposition is analogous to the binomial (series) expansion for 
$\alpha < 0$. After its application, various more Feynman parameters can be integrated using the technique 
described in Section~\ref{sec:32}. In every application the decomposition introduces a number of infinite sums of depth 
one according to the residue theorem. There exist some packages for Mellin--Barnes integrals 
\cite{Czakon:2005rk,Smirnov:2009up,Gluza:2007rt,Gluza:2010rn}, allowing also for numerical checks.
Finally all the produced sums have to be solved using multi--summation methods. Therefore one is advised
to apply this method very carefully. Not all expressions generated by this method can be analytically summed using the
presently know technologies, cf.~\cite{Schneider:2007a,Schneider:2013a}. Sometimes Mellin--$N$ space techniques may
lead to elliptic structures, while $x$--space techniques do not, cf.~\cite{Ablinger:2017xml}, and sum--representations 
have to be cast back into definite integral representations first.
%%%%%%%%%%%%%%%%%%%%%%%%%%%%%%%%%%%%%%%%%%%%%%%%%%%%%%%%%%%%%%%%%%%%%%%%%%%%%%%%%%%%%%%%%%%%%%%
\subsection{Hyperlogarithms}
\label{sec:34}
%%%%%%%%%%%%%%%%%%%%%%%%%%%%%%%%%%%%%%%%%%%%%%%%%%%%%%%%%%%%%%%%%%%%%%%%%%%%%%%%%%%%%%%%%%%%%%%

\vspace{1mm}
\noindent
In a wide class of cases Feynman integrals can be represented by combinations of Kummer--Poincar\'e 
integrals \cite{KUMMER,POINCARE,LADAN,CHEN,GONCHAROV} for (a part) of their expansion coefficients 
in $\ep$. Let us assume one can isolate these terms, see \cite{vonManteuffel:2014qoa}, and forms
a corresponding finite multi--integral. The method of hyperlogarithms \cite{Brown:2008um} has originally 
intended to reorganize these integrals such that one can find a sequence of integrations being
linear in the Feynman parameter on hand. If this is the case the result is given in terms of  
Kummer--Poincar\'e integrals. For a corresponding implementation see \cite{Panzer:2014caa}. The method
has first been applied to the usual massless Feynman integrals. A generalization for massive integrals
also containing local operator insertions has been given in \cite{Ablinger:2014yaa}, with an implementation
in \cite{Wissbrock:2015faa}. Here also certain non--linear Feynman parameter structures, breaking 
multi--linearity, 
could be integrated.
%%%%%%%%%%%%%%%%%%%%%%%%%%%%%%%%%%%%%%%%%%%%%%%%%%%%%%%%%%%%%%%%%%%%%%%%%%%%%%%%%%%%%%%%%%%%%%%
\subsection{The Method of Differential Equations}
\label{sec:35}
%%%%%%%%%%%%%%%%%%%%%%%%%%%%%%%%%%%%%%%%%%%%%%%%%%%%%%%%%%%%%%%%%%%%%%%%%%%%%%%%%%%%%%%%%%%%%%%

\vspace{1mm}
\noindent
In single--scale processes systems of ordinary differential equations for the master integrals are 
naturally obtained by the IBP--relations differentiating for a parameter $x$.\footnote{Correspondingly, in the 
case of more parameters, partial differential equation systems are obtained.} The master integrals may 
then be calculated by solving these systems under given physical boundary conditions,
\cite{Kotikov:1990kg,Bern:1992em,Remiddi:1997ny,Gehrmann:1999as}. One considers the system
%-------------------------------------------------------------------------------------------------------------
\begin{eqnarray}
\label{eq:DEQ1}
\frac{d}{dx}
\left(
\begin{array}{c}
f_1\\ \vdots \\ f_n\end{array}\right)
= \left(\begin{array}{ccc}
A_{11} & \hdots & A_{1,n}
\\
\vdots &  & \vdots
\\
A_{n1} & \hdots & A_{n,n}
\end{array} \right)
\left(\begin{array}{c}f_1\\ \vdots \\ f_n\end{array}\right)
+ \left(\begin{array}{c}g_1\\ \vdots \\ g_n\end{array}\right),
\end{eqnarray}
%-------------------------------------------------------------------------------------------------------------
which may also be transformed into the scalar differential equation
%-------------------------------------------------------------------------------------------------------------
\begin{eqnarray}
\label{eq:DEQ2}
\sum_{k=0}^n p_{n-k}(x)\frac{d^{n-k}}{dx^{n-k}} f_1(x) = \overline{g}(x),
\end{eqnarray}
%-------------------------------------------------------------------------------------------------------------
with $p_n\neq0$, and $(n-1)$ equations for the remaining solutions, which are fully determined by the solution $f_1(x)$.
In setting up these systems one has to perform the expansion in $\ep$ in parallel in the decoupling.

An important class of differential equations is formed by the first order factorizing systems, after 
applying the decoupling methods \cite{Zuercher:94,Ablinger:2013jta} encoded in {\tt Oresys} \cite{ORESYS}, which appear as 
the 
simplest case.  Eq.~(\ref{eq:DEQ1}) may be transformed into Mellin space, decoupled there and solved using the  
efficient methods of the package {\tt Sigma}, cf.~Ref.~\cite{Ablinger:2015tua}.

The decoupled differential operator of (\ref{eq:DEQ2}) can be written in form of a combination of iterative integrals, 
cf.~Section~\ref{sec:42},
%-------------------------------------------------------------------------------------------------------------
\begin{eqnarray}
\label{eq:DEQ4}
f_1(x) &=& \sum_{k = 1}^{n+1} \gamma_k g_{k}(x),~\gamma_k \in {\mathbb C},\\
g_{k}(x) &=& h_0(x)\int_0^x dy_1 h_{1}(y_1) \int_0^{y_1} dy_2 h_{2}(y_2) ...
\int_0^{y_{k-2}} dy_{k-1} h_{k-1}(y_{k-1})
\nonumber\\ && \times
\int_0^{y_{k-1}} dy_{k} q_k(y_{k})
\end{eqnarray}
%-------------------------------------------------------------------------------------------------------------
with $q_k(x)=0$ for $1\leq k\leq m$. Further, $\gamma_{m+1}=0$ if $\bar{g}(x)=0$
in~\eqref{eq:DEQ2}, and $\gamma_{m+1}=1$ and $q_{m+1}(x)$ being a mild variation of
$\bar{g}(x)$ if $\bar{g}(x)\neq0$. These solutions are d'Alembertian~\cite{Abramov:94} since
the master integrals appearing in quantum field theories obey differential equations with rational
coefficients, the letters $h_i$, which constitute the iterative integrals, have to be algebraic. The 
solution can be computed using the package~\texttt{HarmonicSums}~\cite{Ablinger:2017Mellin}. More generally, also Liouvillian 
solutions~\cite{Singer:81} can be calculated with~\texttt{HarmonicSums} utilizing Kovacic's algorithm~\cite{Kovacic:86}.
This algorithm has been applied in many massive three--loop calculations so far, see also 
\cite{Ablinger:2015tua,Ablinger:2017ptf,Ablinger:2017hst,Ablinger:2018zwz}.

If being transformed to the associated system of difference equations, the same
holds, if this system is also first order factorizing. The solution of the remaining equations are directly obtained
by the first solution. 

In the multi--variate case, the $\varepsilon$--representation of a linear system of partial differential 
equations
%---------------------------------------------------------------------------------------------------------------------
\begin{eqnarray}
\label{eq:DEQ5a}
\partial_m f(\varepsilon, x_n) = A_m(\varepsilon, x_n) f(\varepsilon, x_n)
\end{eqnarray}
%---------------------------------------------------------------------------------------------------------------------
is important, as has been recognized in  Refs.~\cite{Kotikov:2010gf,Henn:2013pwa}, see also \cite{Henn:2014qga}.
The matrices $A_n$ can now be transformed in the non--Abelian case by
%---------------------------------------------------------------------------------------------------------------------
\begin{eqnarray}
A_m' = B^{-1} A_m B - B^{-1}(\partial_m B),
\end{eqnarray}
%---------------------------------------------------------------------------------------------------------------------
see also \cite{NOVIKOV:80,Sakovich:1995}, and  one now intends to find a matrix $B$ to transform 
(\ref{eq:DEQ5a}) into the form
%---------------------------------------------------------------------------------------------------------------------
\begin{eqnarray}
\label{eq:DEQ6}
\partial_m f(\varepsilon, x_n) = \varepsilon A_m(x_n) f(\varepsilon, x_n),
\end{eqnarray}
%---------------------------------------------------------------------------------------------------------------------
if possible. This then allows solutions in terms of iterative integrals. A formalism for the basis change to the
$\varepsilon$--basis has been proposed in \cite{Lee:2014ioa} and implemented in the single--variate case in
\cite{Prausa:2017ltv,Gituliar:2017vzm} and in the multi--variate case in \cite{Meyer:2017joq}. 
%%%%%%%%%%%%%%%%%%%%%%%%%%%%%%%%%%%%%%%%%%%%%%%%%%%%%%%%%%%%%%%%%%%%%%%%%%%%%%%%%%%%%%%%%%%%%%%
\subsection{The Method of Arbitrary Large Moments}
\label{sec:36}
%%%%%%%%%%%%%%%%%%%%%%%%%%%%%%%%%%%%%%%%%%%%%%%%%%%%%%%%%%%%%%%%%%%%%%%%%%%%%%%%%%%%%%%%%%%%%%%

\vspace{1mm}
\noindent
In the case of single--scale problems the corresponding class of Feynman integrals depends 
on a real parameter $x \in [0,1]$, which is given e.g. as the ratio of two Lorentz invariants.
For any power in $\ep$ one would like to find the corresponding function in $x$ analytically.
In a series of cases, cf. e.g. \cite{Ablinger:2014vwa,Ablinger:2014nga,Henn:2016tyf,Ablinger:2018zwz},
one may represent the solution in terms of a formal Taylor series in the variable $x$. The differential 
equations implied by the integration-by-parts method 
\cite{Chetyrkin:1981qh,Laporta:2001dd,Studerus:2009ye,vonManteuffel:2012np,MARSEID} can now be turned
into recurrences using the Taylor series ansatz. In solving the corresponding system one may generate
a large number of Mellin moments for the different projections on the individual color factors 
and multiple zeta values \cite{Blumlein:2009cf}. This is the case independently of the fact that the
corresponding $x$-- or $N$--space solution is given by iterative integrals or iterative--noniterative 
integrals. The corresponding method has been described in Ref.~\cite{Blumlein:2017dxp}.
These moments can then be used as an input to the method described in Section~\ref{sec:37}
to find the associated difference equations. In some applications for single scale massive three--loop 
integrals \cite{Ablinger:2017ptf} 8000 moments could be calculated. This is by far more than possible
using standard methods like {\tt Mincer} \cite{Gorishnii:1989gt}, {\tt MATAD} \cite{Steinhauser:2000ry} 
or {\tt Q2E} \cite{Harlander:1997zb,Seidensticker:1999bb}. Based on this number of moments, the formal 
power series may be used as highly precise semi--analytic numeric representations, in case the 
corresponding series expansion has been performed for the physical quantity to be evaluated.
If analytic continuations are still necessary, the method cannot be applied directly.
%%%%%%%%%%%%%%%%%%%%%%%%%%%%%%%%%%%%%%%%%%%%%%%%%%%%%%%%%%%%%%%%%%%%%%%%%%%%%%%%%%%%%%%%%%%%%%%
\subsection{Guessing One-dimensional Integrals}
\label{sec:37}
%%%%%%%%%%%%%%%%%%%%%%%%%%%%%%%%%%%%%%%%%%%%%%%%%%%%%%%%%%%%%%%%%%%%%%%%%%%%%%%%%%%%%%%%%%%%%%%

\vspace{1mm}
\noindent
As has been described in Section~\ref{sec:36} single--variate multiple Feynman parameter integrals
can be either expanded into formal Taylor series or can be Mellin--transformed
%-------------------------------------------------------------------------------------------------------------
\begin{eqnarray}
\label{eq:MELLIN}
G(N) = \Mvec[f(x)](N) = \int_0^1 dx x^{N-1} f(x).
\end{eqnarray}
%-------------------------------------------------------------------------------------------------------------
In both cases one tries now to find the associated difference
equation \cite{NOERLUND} to the set of moments, e.g. $\{G(2),G(4),....,G(2m)\}, m \in \mathbb{N}$
\cite{Larin:1993vu,Larin:1996wd,Retey:2000nq,Blumlein:2004xt}. 
Indeed such an equation exists in many cases, as e.g. for (massive) operator matrix elements \cite{Bierenbaum:2009mv},
but also for single--scale Wilson coefficients, Ref.~\cite{Vermaseren:2005qc}. If a suitably large number of moments
has been calculated analytically the associated series of rational numbers can now be used as input for the guessing
algorithm \cite{GSAGE}, which is also available in {\tt Sage} \cite{SAGE}, exploiting the fast integer algorithms 
available there. 
The method finally returns the wanted difference equation, and tests it by a larger series of further moments.
This method has been applied in Ref.~\cite{Blumlein:2009tj} to obtain from more than 5000 moments the massless
unpolarized three--loop anomalous dimensions and Wilson coefficients in deep-inelastic scattering
\cite{Vogt:2004mw,Moch:2004pa,Vermaseren:2005qc}.
Recently, the method has been applied {\it ab initio} in the calculation of three--loop splitting functions
\cite{Ablinger:2017tan}
and the massive two-- and three--loop form factor \cite{Ablinger:2018zwz,Ablinger:2018yae}. In the case 
of a 
massive operator matrix 
element 
8000 moments \cite{Ablinger:2017ptf} could be calculated and difference equations
were derived for all
contributing color and $\zeta$-value structures. Analytic solutions can be found using the package {\tt
Sigma} \cite{Schneider:2007a,Schneider:2013a}, provided the problem is solvable in difference field theory.
In other cases at least the first order factorizing parts can be factored off. Other techniques are then needed 
to determine the remainder part of the solution.
%--------------------------------------------------
%%%%%%%%%%%%%%%%%%%%%%%%%%%%%%%%%%%%%%%%%%%%%%%%%%%%%%%%%%%%%%%%%%%%%%%%%%%%%%%%%%%%%%%%%%%%%%%
\subsection{The Almkvist--Zeilberger Algorithm}
\label{sec:38}
%%%%%%%%%%%%%%%%%%%%%%%%%%%%%%%%%%%%%%%%%%%%%%%%%%%%%%%%%%%%%%%%%%%%%%%%%%%%%%%%%%%%%%%%%%%%%%%

\vspace{1mm}
\noindent
Since Feynman parameter integrals, depending on an additional parameter $x$, can be given as integrals 
over $\{x_i|_{1=1}^n\} \in [0,1]^n$, they form the multi--integral $I(x)$, depending also on $\ep$. The 
dependence on the real parameter $x$ may be transformed into one on an integer parameter $N$, see 
Section~\ref{sec:36}. The Almkvist--Zeilberger algorithm \cite{AZ1,AZ2} is providing a method to 
find either an associated differential equation for $I(x)$ or a difference equation for $I(N)$, the 
coefficients of which are either polynomials in $\{x,\ep\}$ or $\{N,\ep\}$, 
%-------------------------------------------------------------------------------------------------------------
\begin{eqnarray}
\sum_{l=0}^m P_l(x,\varepsilon) \frac{d^l}{dx^l} I(x,\varepsilon) &=& N(x,\varepsilon)
\label{eq:AZ1}
\end{eqnarray}
\begin{eqnarray}
\sum_{l=0}^m R_l(N,\varepsilon) I(N+l,\varepsilon) &=& M(N,\varepsilon).
\label{eq:AZ2}
\end{eqnarray}
%-------------------------------------------------------------------------------------------------------------

\noindent
Both equations may be inhomogeneous, where the inhomogeneities emerge as known functions from lower order problems.
An optimized and improved algorithm for the input class of Feynman integrals has
been implemented in the \texttt{MultiIntegrate} package~\cite{Ablinger:2013hcp,Ablinger:2015tua}.
It can either produce homogeneous equations of the form~(\ref{eq:AZ1},\ref{eq:AZ2}) or equations with an 
inhomogeneity formed out of already known functions.
%--------------------------------------------------------------------------------
%%%%%%%%%%%%%%%%%%%%%%%%%%%%%%%%%%%%%%%%%%%%%%%%%%%%%%%%%%%%%%%%%%%%%%%%%%%%%%%%%%%%%%%%%%%%%%%
\subsection{Iterative-Noniterative Integrals and Elliptic Solutions}
\label{sec:39}
%%%%%%%%%%%%%%%%%%%%%%%%%%%%%%%%%%%%%%%%%%%%%%%%%%%%%%%%%%%%%%%%%%%%%%%%%%%%%%%%%%%%%%%%%%%%%%%

\vspace{1mm}
\noindent
Non--first order factorizing systems of differential or difference equations for the master integrals, cf. 
Section~\ref{sec:35}, occur at a certain order in massive Feynman diagram calculations. Well--known
examples for this are the sun--rise integral, 
cf.~e.g.~\cite{Broadhurst:1993mw,Laporta:2004rb,Bloch:2013tra,Adams:2013kgc,Adams:2014vja,Adams:2015gva,Adams:2015ydq},
the kite integral \cite{SABRY,Remiddi:2016gno,Adams:2016xah}, the three--loop QCD--corrections to the 
$\rho$--parameter 
\cite{Ablinger:2017bjx,Grigo:2012ji,Blumlein:2018aeq}, and the three--loop QCD corrections to the massive operator 
matrix element $A_{Qg}$ 
\cite{Ablinger:2017ptf}. After separating the first--order factorizing factors a Heun differential equation 
\cite{HEUN} remains in the case of the $\rho$--parameter. One may write the corresponding solution also using 
$_2F_1$--functions with rational argument \cite{IVH,Ablinger:2017bjx} and rational parameters. 
It is now interesting to see whether these solutions can be expressed in terms of complete elliptic 
integrals, which can be checked algorithmically using the triangle group \cite{TAKEUCHI}.

In the examples mentioned one can find representations in terms of complete elliptic integrals of the first 
and second kind, {\bf K} and {\bf E}, cf.~\cite{TRICOMI,WITWAT}, and the question arises whether an argument 
translation allows for a representation through only {\bf K}. Criteria for this have been given in
\cite{Herfurtner1,Movasati1}. In the case of the three--loop QCD-corrections to
the $\rho$--parameter, however, this is not possible.

The homogeneous solution of the Heun equations are given by $_2F_1$--solutions $\psi_k^{(0)}(x), k=1,2$, at a 
specific rational argument. These integrals cannot be represented such that the variable $x$ just appears in
the boundaries of the integral. The inhomogeneous solution reads 
%-------------------------------------------------------------------------------------------------------------
\begin{eqnarray}
\label{eq:INH}
\psi(x) = 
\psi^{(0)}_1(x)\left[C_1 - \int dx \psi_2^{(0)}(x) \frac{N(x)}{W(x)}\right] + \{1 \rightarrow 2\},
\end{eqnarray}
%-------------------------------------------------------------------------------------------------------------
with $N(x)$ and $W(x)$ the inhomogeneity  and the Wronskian. $C_{1,2}$ are the integration constants. Through
partial integration the ratio $N(x)/W(x)$ can be transformed into an iterative integral. Since  
$\psi_{k}^{(0)}(x)$ cannot be written as iterative integrals, $\psi(x)$ is
obtained as an {\it iterative non--iterative integral} 
\cite{Blumlein:2016a,Ablinger:2017bjx} of the type
%-------------------------------------------------------------------------------------------------------------
\begin{eqnarray}
\label{eq:Hit}
&& \hspace*{-8mm}
\mathbb{H}_{a_1,...,a_{m-1};{a_m,F_m(r(y_m))},a_{m+1},...a_q}(x) =
\nonumber\\ 	
&& \hspace*{-8mm}
\int_0^x dy_1 f_{a_1}(y_1) \int_0^{y_1} dy_2 ... \int_0^{y_{m-1}} dy_m f_{a_m}(y_m) F_m[r(y_m)] 
H_{a_{m+1},...,a_q}(y_m),
\end{eqnarray}
%-------------------------------------------------------------------------------------------------------------
with $r(x)$ a rational function and $F_m$ a non--iterative integral. Usually more than one non--iterative 
integral will appear in (\ref{eq:Hit}). $F_m$  denotes {\it any} non--iterative integral, implying 
a very general representation, cf.~\cite{Ablinger:2017bjx}.\footnote{This 
representation has been used in a more special form also in \cite{Remiddi:2017har} later.}
In Ref.~\cite{Adams:2018yfj} an $\varepsilon$--form for the Feynman 
diagrams of elliptic cases has been found recently. However, transcendental letters contribute here. This is in 
accordance with our earlier finding, Eq.~(\ref{eq:Hit}), which, as well is an iterative integral over all objects 
between the individual iterations and to which now also the non--iterative higher transcendental functions 
$F_m[r(y_m)]$ contribute. One may obtain fast convergent representations of $\mathbb{H}(x)$ by overlapping series 
expansions around $x = x_0$ outside possible singularities, see Ref.~\cite{Ablinger:2017bjx} for details.

Let us return to the elliptic case now. Here one one may transform the kinematic variable $x$ occurring as
${\rm \bf K}(k^2) = {\rm \bf K}(r(x))$ into the variable $q = \exp[i\pi \tau]$ analytically with
%-------------------------------------------------------------------------------------------------------------
\begin{eqnarray}
\label{eq:EL1}
k^2 = r(x) = \frac{\vartheta_2^4(q)}{\vartheta_3^4(q)},
\end{eqnarray}
%-------------------------------------------------------------------------------------------------------------
by applying a 3rd order Legendre--Jacobi transformation, where $\vartheta_l, l=1,...,4$ denote Jacobi's 
$\vartheta$-functions and ${\sf Im}(\tau) > 0$.  In this way Eq.~(\ref{eq:INH}) 
is rewritten in terms of the new variable. The integrands are given by products of meromorphic modular 
forms, cf.~\cite{SERRE,COHST,ONO1}, which can be written as a linear combination of ratios of Dedekind's 
$\eta$-function
%-------------------------------------------------------------------------------------------------------------
\begin{eqnarray}
\label{eq:EL2}
\eta(\tau) = q^{\tfrac{1}{12}} \prod_{k=1}^\infty (1-q^{2k})~.
\end{eqnarray}
%-------------------------------------------------------------------------------------------------------------
Depending on the largest multiplier $k \in \mathbb{N}$, $k_m$, of $\tau$ in the argument of the $\eta$-function, 
the solution transforms under the congruence subgroup $\Gamma_0(k_m)$. One can perform Fourier expansions in $q$ 
around the different cusps of the problem, cf.~\cite{ZUDILIN,BROADH18}. 

In the case that the occurring modular forms are holomorphic, one obtains representations in Eisenstein 
series with character, while in the meromorphic case additional $\eta$--factors in the denominators are present.
In the former case the $q$--integrands can be written in terms of elliptic polylogarithms 
in the representation \cite{Adams:2014vja,Adams:2015gva}
%-------------------------------------------------------------------------------------------------------------
\begin{eqnarray}
\label{eq:EL3}
{\rm ELi}_{n,m}(x,y) = 
\sum_{k=1}^\infty
\sum_{l=1}^\infty \frac{x^k}{k^n} \frac{y^l}{l^m} q^{k l}
\end{eqnarray}
%-------------------------------------------------------------------------------------------------------------
and products thereof, cf.~\cite{Adams:2015gva}. The corresponding $q$--integrals can be directly performed.
The solution (\ref{eq:INH}) usually appears for single master integrals. Other master integrals are obtained
integrating further other letters, so that finally representations by $\mathbb{H}(x)$ occur.
Iterated modular forms, resp. Eisenstein series, have been also discussed recently in 
\cite{Adams:2017ejb,Broedel:2018iwv}.
Efficient numerical calculations of modular forms based on $q$--series were obtained in \cite{Bogner:2017vim}.

For systems which factorize only to 3rd and higher order much less is known.
%%%%%%%%%%%%%%%%%%%%%%%%%%%%%%%%%%%%%%%%%%%%%%%%%%%%%%%%%%%%%%%%%%%%%%%%%%%%%%%%%%%%%%%%%%%%%%%
\subsection{Iterative Integrals of Functions with More Variables}
\label{sec:310}
%%%%%%%%%%%%%%%%%%%%%%%%%%%%%%%%%%%%%%%%%%%%%%%%%%%%%%%%%%%%%%%%%%%%%%%%%%%%%%%%%%%%%%%%%%%%%%%

\vspace{1mm}
\noindent
The occurrence of several masses or additional external non--factorizing scales in higher order loop- 
and phase--space integrals leads in general to rational  and root--valued letters with real parameter
letters in the contributing alphabet, 
cf.~\cite{Ablinger:2017xml,Ablinger:2018brx,Blumlein:2019srk,Blumlein:2019qze,Blumlein:2019zux}.
In the case of the loop integrals one obtains letters of the kind
%------------------------------------------------------------------------------------------------ 
\begin{eqnarray}
\frac{1}{1-x(1-\eta)},~\frac{\sqrt{x(1-x)}}{\eta + x(1-\eta)},~\sqrt{x(1-\eta(1-x)},~~\eta \in [0,1].
\end{eqnarray}
%------------------------------------------------------------------------------------------------ 
The iterative integrals and constants which appeared in \cite{Ablinger:2017xml,Ablinger:2018brx}
could finally be all integrated to harmonic polylogarithms containing complicated arguments, at least up to 
one remaining integration, which allows their straightforward numerical evaluation.

In the case of phase space integrals with more scales, e.g. \cite{Blumlein:2019srk,Blumlein:2019qze}, 
also letters contribute, which may imply incomplete elliptic integrals and iterated structures thereof.
Contrary to the functions obtained in Section~\ref{sec:39} these are still iterative integrals, because
the boundaries of the phase--space integrals are real parameters and not constants. The integrands could 
not by rationalized completely by variable transformations, see also \cite{Besier:2018jen}. Contributing 
letters are e.g.
%------------------------------------------------------------------------------------------------ 
\begin{eqnarray}
\frac{x}{\sqrt{1-x^2}\sqrt{1-k^2x^2}},~
\frac{x}{\sqrt{1-x^2}\sqrt{1-k^2x^2}(k^2(1-x^2(1-z^2))-z^2)},
\end{eqnarray}
%------------------------------------------------------------------------------------------------ 
with $k, z \in [0,1]$. The corresponding iterative integrals are called Kummer--elliptic integrals. 
They are derived using the techniques described in Refs.~\cite{Ablinger:2014bra,RAAB1,GuoRegensburgerRosenkranz}.
%%%%%%%%%%%%%%%%%%%%%%%%%%%%%%%%%%%%%%%%%%%%%%%%%%%%%%%%%%%%%%%%%%%%%%%%%%%%%%%%%%%%%%%%%%%%%%%
\section{A Series of Function Spaces}
\label{sec:4}
%%%%%%%%%%%%%%%%%%%%%%%%%%%%%%%%%%%%%%%%%%%%%%%%%%%%%%%%%%%%%%%%%%%%%%%%%%%%%%%%%%%%%%%%%%%%%%%

\vspace{1mm}\noindent
Intermediary and final results for zero-- and single--scale multi--loop calculations have representations 
by special functions as polynomials over $\mathbb{Q}$. In the case of of zero--scale quantities these 
are special numbers. For single scale quantities one either uses finite nested sum representations
in Mellin $N$--space or iterative integral representations in $x$--space. Here $x$ denotes a Lorentz
invariant ratio of two physical quantities. Both spaces are related to each other by the Mellin transform
(\ref{eq:MELLIN}), where $f(x)$ denotes an iterative integral.
The zero--scale quantities can be obtained e.g. in the limit $N \rightarrow \infty$ of these Mellin transforms
or by the values $f(x=1)$. 
%%%%%%%%%%%%%%%%%%%%%%%%%%%%%%%%%%%%%%%%%%%%%%%%%%%%%%%%%%%%%%%%%%%%%%%%%%%%%%%%%%%%%%%%%%%%%%%
\subsection{Classes of Nested Sums}
\label{sec:41}
%%%%%%%%%%%%%%%%%%%%%%%%%%%%%%%%%%%%%%%%%%%%%%%%%%%%%%%%%%%%%%%%%%%%%%%%%%%%%%%%%%%%%%%%%%%%%%%

\vspace{1mm}
\noindent
The methods described in Section~3 very often lead to finite nested sum representations for which algorithms 
exist \cite{Schneider:2007a,Schneider:2013a} to cast these sums into indefinitely nested sums. They are
given by
%---------------------------------------------------------------------------------------------
\begin{eqnarray}
\label{eq:SUM}
S_{b,\vec{a}}(N) = \sum_{k=1}^N g_b(k) S_{\vec{a}}(n),~~~S_\emptyset = 1, g_c \in \bar{\mathfrak{A}},
\end{eqnarray}
%---------------------------------------------------------------------------------------------
with $\bar{\mathfrak{A}}$ the associated alphabet of functions. The sums obey quasi--shuffle relations 
\cite{HOFFMAN,Blumlein:2003gb}. The simplest structures are the finite
harmonic sums \cite{Vermaseren:1998uu,Blumlein:1998if}, where $g_b(k) = ({\rm sign}(b))^k/k^{|b|},~~b \in \mathbb{N} 
\backslash \{0\}$. A generalization is obtained in the cyclotomic case \cite{Ablinger:2011te}. Here the characteristic
summands are $g_{a,b,c}(k) = (\pm 1)^k/(a k +b)^{c}$, with $a,b,c \in \mathbb{N} \backslash \{0\}$.  Further, the 
generalized harmonic sums have letters
of the type $b^k/k^c$, with $c \in \mathbb{N} \backslash \{0\}, b \neq 0, b \in \mathbb{R}$, \cite{Ablinger:2013cf}.
Another generalization are nested finite binomial and inverse-binomial sums, containing also other sums discussed 
before.
An example is given by
%-------------------------------------------------------------------------------------------------------------
\begin{eqnarray}
&& 
\sum_{i=1}^N \binom{2i}{i} (-2)^i \sum_{j=1}^i \frac{1}{\displaystyle j \binom{2j}{j}}
S_{1,2}\left(\frac{1}{2},1\right)(j)
=
\int_0^1 dx \frac{(-x)^N-1}{x+1}\sqrt{\frac{x}{8-x}}
\nonumber\\
&&
\times \Bigl[\HA_{\sf w_{12},1,0}(x)-2\HA_{\sf w_{13},1,0}(x)
-\zeta_2\left(\HA_{\sf w_{12}}(x)-2\HA_{\sf w_{13}}(x)\right)\Bigr]
\nonumber\\ &&
-\frac{5\zeta_3}{8\sqrt{3}}\int_0^1 dx \frac{(-2x)^N-1}{x+\frac{1}{2}}\sqrt{\frac{x}{4-x}}
+c_1\int_0^1 dx \frac{(-8x)^N-1}{x+\frac{1}{8}}\sqrt{\frac{x}{1-x}},
\end{eqnarray}
%-------------------------------------------------------------------------------------------------------------
with $c_1 \approx 0.10184720\dots$, cf.~\cite{Ablinger:2014bra}. Here the indices ${\sf w_k}$ label specific
letters given in \cite{Ablinger:2014bra}.
Infinite binomial and inverse binomial sums have been considered in \cite{Davydychev:2003mv,Weinzierl:2004bn}.
Given the general structure of (\ref{eq:SUM}) 
many more iterated sums can be envisaged and may still appear in even higher order calculations.
%%%%%%%%%%%%%%%%%%%%%%%%%%%%%%%%%%%%%%%%%%%%%%%%%%%%%%%%%%%%%%%%%%%%%%%%%%%%%%%%%%%%%%%%%%%%%%%
\subsection{Classes of Iterated Integrals}
\label{sec:42}
%%%%%%%%%%%%%%%%%%%%%%%%%%%%%%%%%%%%%%%%%%%%%%%%%%%%%%%%%%%%%%%%%%%%%%%%%%%%%%%%%%%%%%%%%%%%%%%

\vspace{1mm}
\noindent
Iterated integrals have the structure
%---------------------------------------------------------------------------------------------
\begin{eqnarray}
\HA_{b,\vec{a}}(x) = \int_0^x dy f_b(y) \HA_{\vec{a}}(y),~~~\HA_\emptyset = 1, f_c \in \mathfrak{A},
\end{eqnarray}
%---------------------------------------------------------------------------------------------
where $f_c$ are real functions and are the letters of the alphabet $\mathfrak{A}$. Iterated integrals 
obey shuffle relations \cite{REUTENAUER,Blumlein:2003gb} which allows to represent them over a multinomial
basis of fewer terms.

The simplest iterative integral having been considered in quantum field theory are the Nielsen integrals
for the two--letter alphabets $\{1/x,1/(1-x)\}$ or $\{1/x,1/(1+x)\}$ \cite{NIELSEN,KMR70,Kolbig:1983qt,Devoto:1983tc},
covering also the polylogarithms \cite{LEWIN1,LEWIN2,Devoto:1983tc}. This class has later been extended to
the harmonic polylogarithms \cite{Remiddi:1999ew} build over the alphabet $\{1/x,1/(1-x),1/(1+x)\}$. A further extension
is to the real representations of the cyclotomic polylogarithms, with $\{1/x,1/\Phi_k(x)\}$
\cite{Ablinger:2011te}, where $\Phi_k(x)$ denotes the $k$th cyclotomic polynomial. Another extension is given by
Kummer--Poincar\'e iterative integrals over the alphabet $\{1/(x-a_i),~~a_i \in \mathbb{C}\}$, 
\cite{KUMMER,POINCARE,LADAN,CHEN,GONCHAROV}. Properties of these functions have been studied in 
Refs.~\cite{Moch:2001zr,Ablinger:2013cf}. In general one may have also more general denominator polynomials $P(x)$,
which one can factor into
%----------------------------------------------------------------------------------------------
\begin{eqnarray}
P(x) = \prod_{k=1}^n (x - a_k) \prod_{l=1}^m (x^2 + b_l x + c_l),~~a_k, b_l, c_l \in \mathbb{R}
\end{eqnarray}
%----------------------------------------------------------------------------------------------
in real representations. One then performs partial fractioning for $1/P(x)$ and forms iterative integrals 
out of the obtained letters. Further classes are found for square--root valued letters as studied e.g. in 
Ref.~\cite{Ablinger:2014bra}. In multi--scale problems, cf.~e.g. 
\cite{Ablinger:2017xml,Blumlein:2019srk,Blumlein:2019qze,Blumlein:2019zux} and Section~\ref{sec:310}, further 
root--valued letters appear, like also the Kummer--elliptic integrals \cite{Blumlein:2019qze}.
%%%%%%%%%%%%%%%%%%%%%%%%%%%%%%%%%%%%%%%%%%%%%%%%%%%%%%%%%%%%%%%%%%%%%%%%%%%%%%%%%%%%%%%%%%%%%%%
\subsection{Classes of Associated Special Numbers}
\label{sec:43}
%%%%%%%%%%%%%%%%%%%%%%%%%%%%%%%%%%%%%%%%%%%%%%%%%%%%%%%%%%%%%%%%%%%%%%%%%%%%%%%%%%%%%%%%%%%%%%%

\vspace{1mm}
\noindent
For the sums of Section~\ref{sec:41} which are convergent in the limit $N \rightarrow \infty$ and the 
iterated integrals of Section~\ref{sec:42} which can be evaluated at $x=1$ one obtains two sets of special 
numbers. They span the solution spaces for zero--scale quantities and appear as boundary values 
for single--scale problems. Examples for these special numbers are the multiple zeta values 
\cite{Blumlein:2009cf}, associated to the harmonic sums and harmonic polylogarithms, special generalized 
numbers \cite{Ablinger:2013cf} like $\Li_2(1/3)$, associated to generalized sums and to Kummer--Poincar\'e iterated 
integrals, special cyclotomic numbers \cite{Ablinger:2011te} like Catalan's number, special binomial numbers 
\cite{Ablinger:2014bra}, as e.g. arccot($\sqrt{7}$),
and special constants in the elliptic case \cite{Ablinger:2017bjx,Laporta:2017okg}. The latter numbers are given by 
integrals
involving complete elliptic integrals at special rational arguments and related functions.
In general these numbers obey more relations than the finite sums and iterated integrals. One may use the 
PSLQ--method to get a first information on relations between these numbers occurring in a given problem and proof
the conjectured relations afterwards.
%%%%%%%%%%%%%%%%%%%%%%%%%%%%%%%%%%%%%%%%%%%%%%%%%%%%%%%%%%%%%%%%%%%%%%%%%%%%%%%%%%%%%%%%%%%%%%%
\subsection{Numerical Representations}
\label{sec:44}
%%%%%%%%%%%%%%%%%%%%%%%%%%%%%%%%%%%%%%%%%%%%%%%%%%%%%%%%%%%%%%%%%%%%%%%%%%%%%%%%%%%%%%%%%%%%%%%

\vspace{1mm}
\noindent
Physical observables based on single scale quantities can either be represented in Mellin $N$--space
or $x$--space. Representations in Mellin $N$--space allow the exact analytic solution of evolution 
equations \cite{Blumlein:1997em} and scheme-invariant evolution equations can be derived in this way 
\cite{Blumlein:2000wh,Blumlein:2004xs}. The $x$--space representation is then obtained by a single 
numerical integral around the singularities of the respective quantity for $N  \in \mathbb{C}$, cf. 
\cite{Blumlein:1997em}, requiring to know the complex representation of the integrand in $N$--space.
In the case of harmonic sums semi--numerical representations were given in \cite{Blumlein:2000hw,Blumlein:2005jg}. 
Furthermore, it is known that basic harmonic sums, except of $S_1(N)$, which is represented by the Digamma 
function, and its polynomials, have a representation by factorial series \cite{FACT1,FACT2}, which has been used 
in \cite{Blumlein:2009fz,Blumlein:2009ta} for their asymptotic representation, see also \cite{Kotikov:2005gr}.
One uses then the recursion relations, which can be obtained from (\ref{eq:SUM}), to move $N \in \mathbb{C}$ 
from the asymptotic region to the desired point on the integration contour in the analyticity region of the problem.
This can be done for the sums of the type being described in 
Refs.~\cite{Ablinger:2011te,Ablinger:2013cf,Ablinger:2014bra} as well, since also in this case asymptotic expansions
can be provided, at least for certain combinations of sums occurring in the respective physical problem, 
cf.~\cite{Ablinger:2014yaa,Ablinger:2015tua}. In the case that the corresponding relations are not given in tabulated 
form, they can be calculated using the package {\tt HarmonicSums} \cite{Vermaseren:1998uu,Blumlein:1998if,
Ablinger:2014rba, Ablinger:2010kw, Ablinger:2013hcp, Ablinger:2011te, Ablinger:2013cf, 
Ablinger:2014bra,Ablinger:2017Mellin}. Relations for harmonic sums are also implemented in {\tt summer} 
\cite{Vermaseren:1998uu}, and for generalized harmonic sums in {\tt nestedsums} \cite{Weinzierl:2002hv}, {\tt 
Xsummer} \cite{Moch:2005uc}, and {\tt PolyLogTools} \cite{Duhr:2019tlz}.

In other applications one may want to work in $x$--space directly. Here numerical representations are available 
for the Nielsen integrals \cite{KMR70}, the harmonic polylogarithms 
\cite{Gehrmann:2001pz,Vollinga:2004sn,Maitre:2005uu,Ablinger:2017tqs,Ablinger:2018sat}, the Kummer--Poincar\'e 
iterative integrals 
\cite{Vollinga:2004sn}, and the cyclotomic harmonic polylogarithms \cite{Ablinger:2018zwz}. These representations are 
also useful to lower the number of numerical integrations for more general problems, e.g. in the multi--variate case.
The relations for the corresponding quantities are implemented for the harmonic polylogarithms in 
\cite{Remiddi:1999ew,Maitre:2005uu} and for all iterative integrals mentioned, including general iterative integrals, in
the package {\tt HarmonicSums}.
%%%%%%%%%%%%%%%%%%%%%%%%%%%%%%%%%%%%%%%%%%%%%%%%%%%%%%%%%%%%%%%%%%%%%%%%%%%%%%%%%%%%%%%%%%%%%%%
\section{Conclusions}
\label{sec:5}
%%%%%%%%%%%%%%%%%%%%%%%%%%%%%%%%%%%%%%%%%%%%%%%%%%%%%%%%%%%%%%%%%%%%%%%%%%%%%%%%%%%%%%%%%%%%%%%

\vspace{1mm}\noindent
In parallel to the analytic higher--loop calculations in Quantum Field Theory the associated 
mathematical methods have been developed by theoretical physicists and mathematicians since 
the 1950ies. We witness a very fast development since the late 1990ies approaching difficult 
massive problems at two--loop and higher order and massless problems form three loops onward. 
The classical methods of polylogarithms and Nielsen--integrals which were standard means, turned 
out to be not sufficient anymore. Since then more and more special number-- and function 
spaces have been revealed, studied and were brought to flexible practical use in very many 
applications. Moreover, a wide host of analytic integration and summation methods has been 
developed during a very short period. In this way very large physics problems could be solved 
analytically -- a triumph of the exact sciences, also thanks to various groundbreaking methods 
in computer algebra. In this context the goal is to improve the accuracy of the fundamental 
parameters of the Standard Model of the elementary particles further. Within the present 
projects this concerns in particular the relative precision of the strong coupling constant 
$\alpha_s(M_Z^2)$ to less than 1\% and of the $\overline{\rm MS}$ mass of the charm quark 
to better than 1.5~\%.

At even higher loop order and for more separated final state legs, introducing more masses and 
kinematic invariants, one expects further mathematical structures to contribute. Possible 
structures of this kind could be Abel--integrals \cite{NEUMANN} and integrals  related to 
K3--surfaces \cite{BSCH}. More inclusive methods, like the method of differential equations,
can certainly determine the degree of non--factorization of a physical problem. However, one would
like to know in a closer sense the respective analytic solution. Here cutting methods can be of use
since the underlying integrands can be systematically related to the final integral by (various)
Hilbert-transform \cite{Hilbert:1912,KRONIG,KRAMERS}.\footnote{For a recent application to the
one--loop case, see e.g. \cite{Abreu:2017mtm}.} In this way integrand structures are revealed, which are
somewhat hidden in the case of differential equations. This method has been advocated early by M.~Veltman
\cite{Veltman:1963th}, see also \cite{REMI1}.

This process to master highly complex Feynman integrals using analytic methods is of course 
just at the beginning and will develop further given the present and future challenges in the 
field. All of these results put experimental analyses in precision measurements at the high 
energy colliders into the position to analyze the data with much reduced theory errors and 
we will get far closer in our insight into the structure of the micro cosmos to 
reveal its ultimate laws. The interdisciplinary joined effort by mathematicians, theoretical 
and experimental particle physicists and experts in computer algebra makes this possible and 
allows to answer quite a series of fundamental scientific questions of our time.

I would like to give my warmest thanks to Peter Paule for his continuous collaboration and 
support to the DESY--RISC collaboration, starting with our first contacts in 2005, arranged 
by Bruno Buchberger. This scientific symbiosis has produced a large number of methods to 
tackle quite a series of difficult problems since, and is continuing to do so in the future. 
Physics, mathematics, and computer algebra profit from this and reach new horizons, which, 
not at all, could have been imagined. In this way we follow together the motto D.~Hilbert 
has given to us:

\begin{center}
{\sf Wir m\"ussen wissen. Wir werden wissen.}

\vspace*{2mm}
{\sf 
Le but unique de la science, c'est l'honneuer de l'esprit humain.}\footnote{Jacobi to 
Legendre, July 2nd, 1830.} 
\end{center}

\vspace{2ex}
\noindent
{\bf Acknowledgment.}~I would like to thank J.~Ablinger, D.~Broadhurst, A.~De Freitas, 
D.~Kreimer, A.~von Manteuffel, P.~Marquard, S.-O.~Moch, W.L.~van Neerven, P.~Paule, 
C.~Schneider, K.~Sch\"onwald, J.~Vermaseren and S.~Weinzierl for countless fruitful 
discussions. This work was supported by the EU TMR network SAGEX Marie Sk\l{}odowska-Curie 
grant agreement No. 764850 and COST action CA16201: Unraveling new physics at the LHC 
through the precision frontier.

%---------------------------------------------------------------------------
\providecommand{\href}[2]{#2}%---------------------------------------------------------------------------

\end{document}